\documentclass[12pt,a4paper]{article}

\usepackage{amsmath}

\numberwithin{equation}{section}

\newcommand{\nc}{\newcommand}
\nc{\rnc}{\renewcommand}

\rnc{\title}[1]{\Large\mbox{}\\ \mbox{}\\
     \textbf{#1}\bigskip\medskip\\}
\rnc{\author}[1]{\large #1\\ \smallskip}
\nc{\address}[1]{{\narrower\normalsize\it #1\\}\bigskip}

\nc{\Wt}[6]{#1\!\!\left.\left(
\begin{matrix}
#5 & #4\\
#2 & #3
\end{matrix}
\right|#6\right)}

\def\phat{\tilde{p}}
\def\qhat{\tilde{q}}
\nc{\thf}{\vartheta_1}
\nc{\thfr}{\vartheta_4}
\nc{\eps}{\epsilon}
\nc{\vareps}{\varepsilon}
\nc{\veps}{\varepsilon}
\nc{\lam}{\lambda}
\nc{\round}[1]{\left(#1\right)}

\def\bfa{\boldsymbol{a}}
\def\bfb{\boldsymbol{b}}
\def\bfT{\boldsymbol{T}}
\def\bfI{\boldsymbol{I}}
\def\bft{\boldsymbol{t}}

\def\th{\vartheta}
\def\half{\frac{1}{2}}
\long\def\omit#1{}
\def\Re{\mbox{Re}}
\def\Im{\mbox{Im}}

\nc{\mf}[1]{#1}

\DeclareMathOperator{\Imag}{Im}
\DeclareMathOperator{\dn}{dn}

\begin{document}

\begin{center}
\title{Scaling Limit of RSOS Lattice Models\\
and TBA Equations}
\author{
Paul A. Pearce\footnote{E-mail address: \texttt{pap@maths.mu.oz.au}}
and Bernard Nienhuis\footnote{E-mail address: \texttt{nienhuis@phys.uva.nl}}
}
\address{\footnotemark[1]Department of Mathematics and Statistics, University of Melbourne,\\
Parkville, Victoria 3052, Australia\\
and\\
\footnotemark[3]Instituut voor Theoretische Fysica, 
Universiteit van Amsterdam,\\
Valckenierstraat 65, 1018XE Amsterdam, The Netherlands.}

\begin{abstract}
\noindent
We study the scaling limits of the $L$-state Restricted Solid-on-Solid
(RSOS) lattice models and their fusion hierarchies in the off-critical 
regimes. 
Starting with the elliptic functional equations of Kl\"umper and Pearce, 
we derive the
Thermodynamic Bethe Ansatz (TBA) equations  of Zamolodchikov. Although this systematic
approach, in principle, allows TBA equations to be derived for all the excited states we restrict 
our attention here to the largest eigenvalue or groundstate in Regimes~III and IV. In Regime~III the
TBA equations are massive while in Regime~IV there is massless scattering 
describing the renormalization group flow between distinct $A_1^{(1)}$ coset 
conformal field theories. Regimes~I and II, pertaining to $Z_{L-1}$ parafermions, will be treated
in a subsequent paper.
\end{abstract}
\end{center}

\section{Introduction}

Many integrable Quantum Field Theories (QFT)~\cite{ZamInt} can be obtained~\cite{ZamPert} by 
perturbing Conformal Field Theories (CFT)~\cite{BPZ} with respect
to an appropriate scaling operator. Examples include both massive and massless
QFTs. From a microscopic point of view, these theories can be obtained as the
continuum scaling limit of two-dimensional integrable lattice
models~\cite{BaxBook}. A primary tool for the study of such theories is the
Thermodynamic Bethe Ansatz (TBA)~\cite{Yang,ZamTBA}. For the most part, the TBA
analysis has been  restricted to the groundstate~\cite{BazhR90} but the method has
been generalized to treat a few excited states~\cite{Martins,Fendley}. A recent
review of these and related topics is given by Mussardo~\cite{Mussardo}.

In this paper we are interested in a systematic derivation of TBA equations
starting at the level of the microscopic integrable lattice models. The approach
developed here should be applicable to all excitations. Although the methods 
are general, we work here in the  framework of the Restricted
Solid-on-Solid (RSOS) lattice models of  Andrews, Baxter and Forrester~\cite{ABF}
and their fusion  hierarchies~\cite{DJKMO}. The key to the derivation of the TBA
equations is in essence to ``solve" the elliptic TBA functional equations of
Kl\"umper and  Pearce~\cite{KP} in the appropriate scaling regime for the
finite-size  corrections to the transfer matrix eigenvalues. The approach of the present paper
provides an alternative approach to the QFT transfer matrix approach introduced recently by
Bazhanov, Lukyanov and Zamolodchikov~\cite{BLZ}.

The layout of the paper is as follows. We begin, in this section,  by defining
the RSOS lattice models and describing their relation to  perturbed CFTs in the
four distinct integrable regimes. We define the commuting transfer matrices and
state the functional equations of Pearce and Kl\"umper~\cite{KP}.  We also
introduce the notion of the scaling limit and give an overview of the  general
structure of the TBA equations. In Sections~2 and 3 we derive the TBA
equations relevant to Regimes~III and IV respectively. Regimes~I and II will be treated in a
subsequent paper. The massive TBA equations in Regimes~II and III have been derived previously by
Bazhanov and Reshetikhin~\cite{BazhR90}. Their approach uses the Bethe ansatz equations and proceeds
subject to a conjecture on the string solutions. Although conjectured by Zamolodchikov~\cite{ZamTBA},
there appears to be no previous derivation of the massless TBA equations in Regime~IV. In some sense
this is the most interesting regime since it describes the massless renormalization group flow between
distinct coset conformal field theories. We conclude the paper with some discussion of future
work.

\subsection{RSOS Models}

The RSOS$(p,q)$ models are restricted solid-on-solid models in which heights
$a_i$ on the sites $i$ of the square lattice take values in the set 
$\{1,2,3,\dots,L\}$, with $L\ge 3$, subject to the  additional condition that
the values
$a_i$, $a_j$ of heights on  adjacent sites of the lattice satisfy the constraints
\begin{equation} 0\le a_i\le L,\qquad 0\le (a_i-a_j+m)/2\le m,\qquad m < a_i+a_j
< 2L-m+2
\end{equation} where $m=p$ for a horizontal pair and $m=q$ for a vertical pair.
The Boltzmann weights $W^{p,q}(u)$ depend on a \emph{spectral parameter} $u$, a
\emph{nome} or temperature variable $t=\phat^2$ and a
\emph{crossing parameter} $\lam=\pi/(L+1)$. These weights are assumed  to vanish
unless the adjacency constraints are satisfied. There are four  distinct
off-critical physical regimes:
\begin{eqnarray} 
\begin{array}{lcr}
\text{Regime I:}&\qquad   -\pi/2+\lambda\le u \le 0,\qquad& -1<t<0\\ 
\text{Regime II:} & -\pi/2+\lambda\le u \le 0,& 0<t<1\\
\text{Regime III:}& 0\le u \le \lambda,& 0<t<1\\
\text{Regime IV:}&  0\le u \le \lambda, &  -1<t<0
\end{array}
\end{eqnarray}

There are two lines of critical points at $t=0$,  one separating Regimes~I and
II and another separating Regimes~III and IV.  In particular, the RSOS$(p,p)$
models are conformally invariant on these  critical lines. On the Regime~III/IV
critical line the models are described by the coset conformal field
theories~\cite{GKO}
\begin{equation} (A_1^{(1)})_{L-p-1}\oplus (A_1^{(1)})_{p}\supset
(A_1^{(1)})_{L-1}
\end{equation} with central charges
\begin{equation} c=\frac{3p}{p+2}\left(1-\frac{2(p+2)}{(L+1)(L+1-p)}\right).
\end{equation} On the Regime~I/II critical line the models are described by
$Z_{L-1}$ parafermion theories~\cite{FatZam} with central charges
\begin{equation} c=\frac{2(L-2)}{L+1}.
\end{equation}

The face weights of the RSOS$(p,q)$ models, with $\max[p,q]\le L-1$, are 
constructed by fusing blocks of $p\times q$ elementary faces together. The
elementary faces with $(p,q)=(1,1)$ are given by the ABF models. In this case 
the adjacency condition reduces to
\begin{equation} 1\le a_i,a_j\le L,\qquad a_i-a_j=\pm 1.
\end{equation} Explicitly, the non-zero face weights $W^{1,1}(u)=W(u)$ are given
by
\begin{eqnarray}
\Wt{W}{a}{a\mp1}{a}{a\pm1}u&=&\frac{\thf(\lam-u)}{\thf(\lam)}\\
\Wt{W}{a\mp1}{a}{a\pm1}{a}u&=&\frac{g_{a\pm1}}{g_{a\mp1}}
\round{\frac{\thf((a-1)\lam)\thf((a+1)\lam)}{\thf^2(a\lam)}}^{1/2}\,
\frac{\thf(u)}{\thf(\lam)}\\
\Wt{W}{a\pm1}{a}{a\pm1}{a}u&=&\frac{\thf(a\lam\pm u)}{\thf(a\lam)}
\end{eqnarray} where $\thf(u)=\thf(u,\phat)$ and the $g_a$ are arbitrary
$u$-independent gauge factors that cancel out  on a periodic lattice and do not
effect the eigenvalues of the transfer  matrices.  In this paper we use standard
elliptic theta functions as given in  Gradshteyn and Ryzhik~\cite{GR}
\begin{eqnarray}
\thf(u,\phat)&=&2\phat^{1/4}\sin u
\prod_{n=1}^\infty (1-2\phat^{2n}\cos 2u+\phat^{4n})(1-\phat^{2n})\\
\th_2(u,\phat)&=&2\phat^{1/4}\cos u
\prod_{n=1}^\infty (1+2\phat^{2n}\cos 2u+\phat^{4n})(1-\phat^{2n})\\
\th_3(u,\phat)&=&\prod_{n=1}^\infty (1+2\phat^{2n-1}\cos 2u+\phat^{2(2n-1)})
(1-\phat^{2n})\\
\thfr(u,\phat)&=&\prod_{n=1}^\infty (1-2\phat^{2n-1}\cos 2u+\phat^{2(2n-1)})
(1-\phat^{2n}).
\end{eqnarray}   We express the nome
$\phat$ in terms of a real parameter $\veps>0$ by
\begin{equation}
\phat=\begin{cases} e^{-\pi\veps} & \text{Regimes II and III},\\ ie^{-\pi\veps}
& \text{Regimes I and IV},
\end{cases}
\label{nome}
\end{equation} so that $t=\phat^2=\pm\exp(-2\pi\veps)$. In particular, the
$\thf$ functions satisfy the quasiperiodicity  properties
\begin{gather}
\thf(u+\pi,\phat)=-\thf(u,\phat)\\
\thf(u-i\ln \phat,\phat)=-\phat^{-1}e^{-2iu}\thf(u,\phat).
\end{gather}

To define the RSOS$(p,q)$ face weights it is convenient to fix a particular gauge
\begin{equation} g_a=(-1)^{a/2}\sqrt{\thf(a\lambda)}.
\end{equation} With this choice of gauge the RSOS$(p,q)$ face weights are given
by
\begin{equation}
\Wt{W^{p,q}}{a_1}{b_1}{b_{q+1}}{a_{q+1}}u
=\prod_{k=0}^{q-2}s_k^p(u)^{-1}\sum_{a_1,\ldots,a_q}
\prod_{k=1}^q \Wt{W^{p,1}}{a_k}{b_k}{b_{k+1}}{a_{k+1}}{u+(k-1)\lambda}
\end{equation} independent of the values of the edge spins $b_2,\ldots,b_q$. Here
\begin{equation}
s_q^p=\prod_{j=0}^{p-1}\frac{\thf(u+(q-j)\lambda)}{\thf(\lambda)}
\end{equation} and the $p\times 1$ face weights are given in turn by
\begin{equation}
\Wt{W^{p,1}}{a_1}{a_{p+1}}{b_{p+1}}{b_1}u=\sum_{a_2,\ldots,a_p}
\prod_{k=1}^p \Wt{W}{a_k}{a_{k+1}}{b_{k+1}}{b_k}{u+(k-p)\lambda}
\end{equation} independent of the edge spins $b_2,\ldots,b_p$.

\subsection{Transfer Matrices and Functional Equations}

Consider the RSOS$(p,q)$ models and suppose that $\bfa$ and $\bfb$ are allowed
height configurations for two consecutive rows of an $N$ column lattice with
periodic boundary conditions. Then the elements of the row transfer matrices of
the RSOS$(p,q)$ models are given by
\begin{equation}
\langle\bfa|\bfT^{p,q}(u)|\bfb\rangle
=\prod_{j=1}^N\Wt{W^{p,q}}{a_j}{a_{j+1}}{b_{j+1}}{b_j}u
\end{equation} where $a_{N+1}=a_1$ and $b_{N+1}=b_1$. The Yang-Baxter equations
\begin{eqnarray} &&\sum_g
\Wt{W^{p,q}}abgfu\Wt{W^{p,s}}fgde{u+v}\Wt{W^{q,s}}gbcdv\nonumber\\ &&\qquad
=\sum_g \Wt{W^{q,s}}fagev\Wt{W^{p,s}}abcg{u+v}\Wt{W^{p,q}}gcdeu
\end{eqnarray} imply that, for fixed $p$, these transfer matrices belong to
commuting families
\begin{equation}
\bfT^{p,q}(u)\bfT^{p,q'}(v)=\bfT^{p,q'}(v)\bfT^{p,q}(u).
\end{equation}

Starting with the fusion hierarchy of Bazhanov and Reshetikhin~\cite{BazhR},
Kl\"umper and Pearce have shown that the commuting transfer matrices satisfy the
inversion identity hierarchy
\begin{equation}
\bfT_0^{p,q}\bfT_1^{p,q}=f_{-1}^p f_q^p\bfI+\bfT_0^{p,q+1}\bfT_1^{p,q-1},
\quad  1\le q\le L-1
\label{invHier}
\end{equation} where
\begin{equation}
\bfT_k^{p,q}=\bfT^{p,q}(u+k\lambda),\qquad f_q^p=s_q^p(u)^N.
\end{equation} These functional equations close with
\begin{equation}
\bfT_0^{p,-1}=0,\quad \bfT_0^{p,0}=f_{-1}^p\bfI,\quad \bfT_0^{p,L}=0.
\end{equation} If we further define
\begin{equation}
\bft_0^{p,q}=\frac{\bfT_0^{p,q-1}\bfT_1^{p,q+1}}{f_{-1}^p f_q^p},
\quad 1\le q\le L-2
\end{equation} then Kl\"umper and Pearce~\cite{KP} have also shown that these
equations can  be recast in the form of TBA functional equations
\begin{equation}
\bft_0^{p,q}\bft_1^{p,q}=(\bfI+\bft_1^{p,q-1})(\bfI+\bft_0^{p,q+1}),
\quad 1\le q\le L-2
\label{TBAfunct}
\end{equation} where
\begin{equation}
\bft_0^{p,0}=\bft_0^{p,L-1}=0.
\end{equation}

From quasiperiodicity, it follows that
\begin{equation}
\frac{\bfT_0^{p,q}}{f_{(q-2)/2}^p}=\text{doubly periodic},
\qquad \bft_0^{p,q}=\text{doubly periodic}
\end{equation} with period rectangle in the complex $u$ plane given by
\begin{equation}
\text{period rectangle}=
\begin{cases}
\pi\times \pi i\veps,&\text{Regimes~II and III}\\
\pi\times 2\pi i\veps,&\text{Regimes~I and IV.}
\end{cases}
\end{equation} In Regimes~I and IV there is an additional symmetry within the
period rectangle
\begin{equation}
\bft^{p,q}(u\pm\pi/2+\pi i\veps)=\bft^{p,q}(u).
\end{equation} These statements apply to the entries of the transfer matrices
and to their eigenvalues.

At the critical point with $t=0$ and in the braid limit
\begin{equation}
\lim_{u\to \pm i\infty}t^{p,q}(u)=t_\infty^{p,q}
\end{equation} the TBA functional equations reduce to a simple recursion
relations which can be solved immediately to give
\begin{equation}
t_\infty^{p,q}=\frac{\sin(q\phi)\sin((q+2)\phi)}{\sin^2\phi},\qquad
1+t_\infty^{p,q}=\frac{\sin^2((q+1)\phi)}{\sin^2\phi}
\end{equation} with the quantization
\begin{equation}
\phi=\frac{m_j\pi}{L+1},\qquad m_j=1,2,\ldots,L.
\end{equation}

\subsection{Scaling Limit and Perturbed CFTs}

Let $a$ denote the lattice spacing. Then the (continuum) scaling limit of the
RSOS models  is given by
\begin{equation} N\to\infty,\quad a\to 0,\quad t\to 0,\quad
\mu=N|t|^{\nu}=\text{fixed}
\label{scaling}
\end{equation} where
\begin{equation}
\nu=
\begin{cases}
\displaystyle{\frac{L+1}{4}},&\text{Regimes~III and IV}\\
\\
\displaystyle{
\frac{L+1}{2(L-1)}},&\text{Regimes~I and II.}
\end{cases}
\label{scaling2}
\end{equation} Here $\nu$ is the critical exponent associated with the
correlation length of the RSOS$(p,p)$ model which is in fact independent of the
fusion level $p$. The exponent $\nu$ is related to the usual specific heat
exponent $\alpha$ by  the hyperscaling relation $2-\alpha=d\nu$. Strictly
speaking, this relation breaks down in Regime~IV for which $\nu'=2\nu$. It is
this fact that leads to Regime~IV being massless. Also, in some cases for $p>1$,
the free energy is analytic~\cite{DJKMO} so $\alpha$ is not defined. We ignore
these problems and scale precisely as indicated in \eqref{scaling} and
\eqref{scaling2}.

In the scaling limit we introduce variables
\begin{equation} R=\lim_{N\to\infty,\ a\to 0} Na,\qquad m=\lim_{t\to 0,\ a\to 0}
\frac{4t^\nu}{a}
\end{equation} where $R$ is the radial distance and $m$ is a mass. It follows
immediately from  these definitions that
\begin{equation} 4\mu=mR.
\end{equation}

For the RSOS$(p,p)$ models in Regimes~III and IV, the off-critical  elliptic
solution corresponds to perturbation away from the conformal critical point by
the  operator $\Phi$ with conformal weights
\begin{equation}
\Delta=\overline{\Delta}=\frac{L-1}{L+1}.
\end{equation} For the RSOS$(1,1)$ models this is the thermal operator
$\Phi=\Phi_{1,3}$.

\subsection{Zamolodchikov's TBA Equations}

The Thermodynamic Bethe Ansatz (TBA) equations describe the scattering of 
$n$ types of relativistic particles on a circle of radius $R$. Here we will be
interested in the TBA equations introduced by Zamoldchikov~\cite{ZamTBA}. All of
these take the form of a system of coupled nonlinear integral equations
\begin{equation}
\epsilon_i(\theta)=m_i R\nu_i(\theta)-\sum_{j=1}^n \frac{1}{2\pi}
\int_{-\infty}^\infty d\theta'\,\Phi_{ij}(\theta-\theta')
\log(1+e^{-\epsilon_j(\theta')})
\label{TBA}
\end{equation} with the Casimir energy
\begin{equation} E(R)=-\sum_{j=1}^n \frac{m_j}{2\pi}
\int_{-\infty}^\infty d\theta\,\nu_j(\theta)\log(1+e^{-\epsilon_j(\theta)}).
\end{equation} Here $m_i$ are the particle masses, 
$\theta$, $\theta'$ are rapidities,
$\epsilon_i(\theta)$ are quasiparticle pseudo-energies,
$\nu_i(\theta)$ are energy functions and the symmetric kernel satisfies 
$\Phi_{ij}(\theta)=\Phi_{ji}(\theta)$. For diagonal scattering, which ensues in
Regimes~I and II but not  Regimes~III and IV, the kernel is related to the $S$
matrix by
\begin{equation}
\Phi_{ij}=-i\frac{\partial}{\partial\theta} \log S_{ij}(\theta).
\end{equation} Typically, the energy functions are given by 
$\nu(\theta)=\frac{1}{2}e^{\pm\theta}$ or $\nu(\theta)=\cosh\theta$.

In the scaling region, the finite-size corrections to the largest eigenvalue
$T(u)$ are related to  the Casimir energy by
\begin{equation} -\log T(u)= N f(u)+\frac{R\sin\vartheta}{N}\,E(R)+o(\frac{1}{N})
\end{equation} where $f(u)$ is the bulk free energy and the anisotropy angle is
given by
\begin{equation}
\vartheta=
\begin{cases} (L+1)u,&\text{Regimes III and IV}\\
-\displaystyle{\frac{2(L+1)u}{L-1}},&\text{Regimes I and II.}
\end{cases}
\end{equation} Despite the appearance of $1/N$ corrections, the system is not in
general  conformally invariant. The system, however, will be conformal at
critical points. Such critical points can occur in the infrared limit
($R\to\infty$ or $\mu\to\infty$) in addition to the critical point in the
ultraviolet limit ($R\to 0$ or $\mu\to 0$) with
\begin{equation} R E(R)\to -\pi c/6
\end{equation} where $c$ is the central charge.

The scaling behaviour is manifest in the behaviour of the zeros of $T(u)$ in the
complex $u$ plane. The imaginary period of $T(u)$ is either $2\pi\veps$ or
$4\pi\veps$ where 
$\pi\veps=2(\log N-\log\mu)/(L+1)$. For large $N$
and finite nonzero $\mu$, the scaling behaviour occurs in the two regions 
\begin{equation}
\Im(\vartheta)= \pm(\theta+\log N-\log\mu)
\end{equation}   where the rapidity $\theta$ is independent of $N$. This
equation relates the rapidity $\theta$ to the spectral parameter $u$ in these
two scaling regions.

\section{Regime III}

In this section we derive the TBA equations in Regime~III. The strategy is to ``solve"
the TBA functional equations in the scaling limit for the finite-size  corrections
to the eigenvalues $t^{p,q}(u)$ of the transfer matrices 
$\bft^{p,q}(u)$. In these equations the dependence on the temperature $t$ enters
only in the combination $t^\nu$ which we replace with $\mu/N$. Regarding $\mu$ as
fixed there only remains a dependence on $N$, but since these functional equations
are exact for finite $N$ they contain all the information required to calculate the
finite size corrections. To do this we follow closely Kl\"umper and
Pearce~\cite{KP}.

\subsection{Integral Equations and Finite-Size Corrections}

As in Kl\"umper and Pearce we consider the analyticity strips of
$t^{p,q}(u)$:
\begin{equation}
\frac{p-q-2}{2}\lambda<\Re(u)<\frac{p-q+2}{2}\lambda.
\end{equation} In Regime~III the leading contributions to the eigenvalues
$t^{p,q}(u)$ in its analyticity strip are given for large $N$ by
\begin{equation} t^{p,q}(u)=
\begin{cases} t_{\text{const}}^{p,q}\displaystyle{
\left[i\frac{\th_1((L+1)u/2,t^\nu)}{\th_2((L+1)u/2,t^\nu)}\right]^N},& q=p\\
\\ t_{\text{const}}^{p,q},& p \neq q
\end{cases}
\end{equation} 
where the constants are given by
\begin{equation} t_{\text{const}}^{p,q}=
\begin{cases}
\displaystyle{
\frac{\sin(q\sigma)\sin((q+2)\sigma)}{\sin^2\sigma}},& 1\le q\le p-1\\
\\ 4\cos\sigma\cos\tau,& q=p\\
\\
\displaystyle{
\frac{\sin((q-p)\tau)\sin((q-p+2)\tau)}{\sin^2\tau}},& p+1\le q\le L-2
\end{cases}
\end{equation} and 
\begin{eqnarray}
\sigma&=&\frac{m_j'\pi}{p+2},\qquad\qquad m_j'=1,2,\ldots,p+1\\
\tau&=&\frac{m_j''\pi}{L+1-p},\qquad m_j''=1,2,\ldots,L-p.
\end{eqnarray} For the largest eigenvalue, we have
\begin{equation}
\phi=\frac{\pi}{L+1},\qquad \sigma=\frac{\pi}{p+2},
\qquad \tau=\frac{\pi}{L+1-p}.
\end{equation} 
Other choices will be needed for the excited states. Note that the
leading contributions to the eigenvalues $t^{p,q}(u)$ give the correct periodicity, have the
required zero and pole structure and that the constants are such that
the functional equations \eqref{TBAfunct} are satisfied including in the scaling
limit.

We introduce functions of a real variable by restricting the eigenvalue functions
to certain lines in the complex $u$ plane
\begin{equation}
\mf{a}^q(x)=t^{p,q}\left(\frac{i}{L+1}x+\frac{p-q}{2}\lambda\right),\qquad
\mf{A}^q(x)=1+\mf{a}^q(x)
\end{equation} Here and the sequel we will usually suppress the dependence on $p$.
Let us also introduce finite-size corrections terms $\ell^q(x)$ by writing
\begin{equation}
\mf{a}^q(x)=\mf{e}^q(x)\ell^q(x)
\end{equation} with
\begin{equation}
\mf{e}^q(x)=
\begin{cases} 
\displaystyle{
\left[i\frac{\th_1(ix/2,t^\nu)}{\th_2(ix/2,t^\nu)}\right]^N},&q=p\\
1,&q\ne p
\end{cases}
\end{equation} 
The TBA functional equations then take the form
\begin{equation}
\ell^q(x-\pi i/2)\ell^q(x+\pi i/2)=\mf{A}^{q-1}(x)\mf{A}^{q+1}(x).
\label{xInvHier}
\end{equation} The functions $\ell^q(x)$ and $\mf{A}^q(x)$ are analytic and nonzero
in the strip $-\pi<\Imag(x)<\pi$ with period $(L+1)\pi\vareps$. We represent their 
logarithmic derivatives by Fourier series
\begin{gather} [\log\ell^q(x)]'=\sum_{k=-\infty}^\infty  L_k^q\, e^{\frac{2
ikx}{(L+1)\vareps}}\\ 
L_k^q=\frac{1}{(L+1)\pi\vareps}
\int_{-(L+1)\pi\vareps/2}^{(L+1)\pi\vareps/2} [\log\ell^q(x)]'\,
e^{-\frac{2ikx}{(L+1)\vareps}}\,dx
\end{gather} with similar equations relating the Fourier coefficients $A_k^q$ to
$\mf{A}^q(x)$.

So taking the logarithmic derivative of \eqref{xInvHier} and equating Fourier 
coefficients gives
\begin{equation}
\left[e^{\frac{k\pi}{(L+1)\vareps}}+
      e^{-\frac{k\pi}{(L+1)\vareps}}\right]L_k^q =A_k^{q-1}+A_k^{q+1}
\label{fourierEqn}
\end{equation} and hence
\begin{equation} [\log\ell^q(x)]'=\sum_{k=-\infty}^\infty
\frac{e^{\frac{2ikx}{(L+1)\vareps}}} {e^{\frac{k\pi}{(L+1)\vareps}}+
      e^{-\frac{k\pi}{(L+1)\vareps}}}(A_k^{q-1}+A_k^{q+1}).
\end{equation} Substituting the integral expression for $A_k^q(x)$ and evaluating
the sum on $k$ gives the nonlinear integral equation
\begin{equation} [\log\ell^q(x)]'=
\int_{-(L+1)\pi\vareps/2}^{(L+1)\pi\vareps/2} dy\, \mf{k}(x-y)
\left\{[\log\mf{A}^{q-1}(y)]'+[\log\mf{A}^{q+1}(y)]'\right\}.
\end{equation} After integration this yields
\begin{equation}
\log\mf{a}^q=\log\mf{e}^q+\mf{k}*\log\mf{A}^{q-1}+\mf{k}*\log\mf{A}^{q+1} +\mf{D}^q
\label{TBAIII}
\end{equation} where $*$ denotes convolution (over the given interval) and
$\mf{D}^q$ are integration constants. 

The kernel is given in terms of standard
Jacobian elliptic functions and theta functions by
\begin{eqnarray}
\mf{k}(x)&=&\frac{1}{(L+1)\pi\vareps}\sum_{k=-\infty}^\infty
\frac{e^{\frac{2ikx}{(L+1)\vareps}}} {e^{\frac{k\pi}{(L+1)\vareps}}+
      e^{-\frac{k\pi}{(L+1)\vareps}}}\nonumber\\
&=&\frac{I'}{\pi^2}\dn\left(\frac{2I'x}{\pi},\qhat\right)
=\frac{1}{2\pi}\frac{\th_1'(0,\qhat')}{\th_4(0,\qhat')}
\frac{\th_3(ix,\qhat')}{\th_2(ix,\qhat')}\\
&=&\frac{\th_2(0,\qhat')\th_3(0,\qhat')\th_3(ix,\qhat')}{
2\pi\,\th_2(ix,\qhat')}\nonumber
\end{eqnarray} where $I$ and $I'$ respectively denote the complete elliptic
integrals of nomes
\begin{equation}
\qhat=e^{-\frac{\pi}{(L+1)\veps}},\qquad
\qhat'=e^{-\pi(L+1)\veps}=\frac{\mu^2}{N^2}.
\end{equation} In the critical limit, when $\veps\to\infty$ and $\qhat'\to 0$, the
kernel simplifies to
\begin{equation}
\lim_{\veps\to\infty}\mf{k}(x)=\hat{k}(x)=\frac{1}{2\pi\cosh x}.
\end{equation}

To evaluate the constants $D^q$ we set
\begin{eqnarray}
\langle a\rangle=\frac{1}{(L+1)\pi\vareps}
\int_0^{(L+1)\pi\vareps}
a(x)\,dx
\end{eqnarray} 
Then integrating \eqref{TBAIII} and using the result
\begin{equation}
\int_0^{(L+1)\pi\vareps} \mf{k}(x)\,dx=\half
\end{equation} 
we find that
\begin{equation}
\langle\log\ell^q\rangle=
\half\left(\langle A^{q-1}\rangle+\langle A^{q-1}\rangle\right)+D^q.
\end{equation} 
Comparing with \eqref{xInvHier} we see that, for each $q$,
$D^q$ is a multiple of $\pi i$ related to the branches of the logarithms and winding. In particular,
for the largest eigenvalue, there is no winding and $D^q=0$. Alternatively, we could work
directly with the Fourier series of the logarithms rather than the logarithmic derivatives. In
this case we would need to allow for multiples of $2\pi i$ on the right side of \eqref{fourierEqn}
with the same end result.

Let us factor the eigenvalues $T^{p,q}(u)$ into bulk and finite-size  correction
terms
\begin{equation} T^{p,q}(u)=T^{p,q}_{\text{bulk}}(u)T^{p,q}_{\text{finite}}(u)
\end{equation} then as in Kl\"umper and Pearce
\begin{equation} T^{p,q}_{\text{bulk}}(u)T^{p,q}_{\text{bulk}}(u+\lambda)
=f(u-\lambda)f(u+q\lambda)
\end{equation} and
\begin{equation} T^{p,q}_{\text{finite}}(u)T^{p,q}_{\text{finite}}(u+\lambda)
=1+t^{p,q}(u)\label{FinInvIII}
\end{equation} where now $T^{p,q}_{\text{finite}}(u)$ is a doubly periodic function
of $u$. It is again useful to introduce functions of a real variable by restricting
the eigenvalue function to certain lines
\begin{equation}
\mf{b}^q(x)=
T^{p,q}_{\text{finite}}\left(\frac{i}{L+1}x+\frac{p-q+1}{2}\lambda\right)
\end{equation} so this relation becomes
\begin{equation}
\mf{b}^q(x-\pi i/2)\mf{b}^q(x+\pi i/2)=\mf{A}^q(x).
\label{xFinInvIII}
\end{equation}

Since $\mf{b}^q$ and $\mf{A}^q$ are non-zero and periodic we can solve as before by
introducing the Fourier series of the logarithmic derivatives to obtain
\begin{equation}
\log\mf{b}^q=\mf{k}*\log\mf{A}^q+\mf{C}^q\label{FinSizeIII}
\end{equation} 
where $\mf{C^q}$ are constants and the convolution is over the finite
interval. Explicitly, the finite-size $1/N$ corrections \eqref{FinSizeIII} for
$q=p$ are
\begin{equation}
\log\mf{b}^p(x)=\int_{-(L+1)\pi\vareps/2}^{(L+1)\pi\vareps/2}dy\;
\mf{k}(x-y)\log(1+\mf{a}^p(y))+\mf{C}^p\label{FinSizeIntIII}
\end{equation} 

We evaluate the constant as before
by integrating \eqref{FinSizeIntIII} to obtain
\begin{equation}
\langle b^p\rangle=\half\langle A^p\rangle+C^p.
\end{equation} 
Comparing with \eqref{xFinInvIII} we see that, for each $p$,
$\mf{C}^p$ is a multiple of $\pi i$ so it does not contribute to the $1/N$ corrections.

\subsection{TBA Equations and Casimir Energy}

So far our integral equations are completely general and are exact for  finite $N$.
They could be used,  for example, to study the exponentially small corrections to
the eigenvalues off-criticality. In this subsection, however, we specialize the
form of these  equations appropriate to the scaling limit. 

The system size $N$ only enters the integral equations through the function
$\mf{e}^p(x)$. This function is periodic in $x$ with period 
$2(\log N-\log\mu)$. For $N$ large the function is exponentially small in
$N$ except in the two scaling regions when $x$ is of the order of $\log N$ or
$-\log N$.  Let us set
\begin{equation} 
z^2=e^{-(x+\log N)}=\frac{e^{-x}}{N},\qquad t^\nu=\frac{\mu}{N}.
\end{equation} 
Then in these scaling regions we find
\begin{eqnarray}
\log \hat{e}_{\pm}^p(x)&=&\lim_{N\to\infty}\log\mf{e}^p(\pm(x+\log N))\nonumber\\
&=&\lim_{N\to\infty} \log\left[
\left(\frac{1-z^2}{1+z^2}\right)^N
\left(\frac{1-t^{2\nu} z^2}{1+t^{2\nu} z^2}\right)^N
\left(\frac{1-t^{2\nu} z^{-2}}{1+t^{2\nu} z^{-2}}\right)^N \ldots
\right]\nonumber\\ &=&-2(e^{-x}+\mu^2e^x).
\end{eqnarray}

We assume that the functions $\mf{a}^q$ and $\mf{A}^q$ scale similarly and set
\begin{eqnarray} 
\hat{a}_{\pm}^q(x)&=&\lim_{N\to\infty}\mf{a}^p(\pm(x+\log N))\\
\hat{A}_{\pm}^q(x)&=&\lim_{N\to\infty}\mf{A}^p(\pm(x+\log N)=1+\hat{a}_{\pm}^q(x).
\end{eqnarray} 
Scaling $x$ and $y$ in the same way, the integral equations
\eqref{TBAIII}  take the following simplified form in the scaling limit
\begin{equation}
\log \hat{a}^q=\log \hat{e}^q+\hat{k}*\log \hat{A}^{q-1}+\hat{k}*\log \hat{A}^{q+1}+D^q
\end{equation} where we suppress the subscripts $\pm$ and
\begin{equation}
\log \hat{e}^q(x)=\begin{cases} 0,&q\ne p\\ -2(e^{-x}+\mu^2e^x),&q=p.
\end{cases}
\end{equation} Clearly, this reduces back to the case of Kl\"umper and Pearce when
$\mu\to 0$. Indeed, the modified equation is obtained just by replacing $-2e^{-x}$
by $-2(e^{-x}+\mu^2e^x)$ in $\hat{e}^p(x)$.

If we now set
\begin{equation}
\theta=x+\log\mu
\end{equation} and
\begin{equation} \hat{a}^q(x)=\hat{a}^q(\theta-\log\mu)=e^{-\epsilon_q(\theta)}
\end{equation} then $-2(e^{-x}+\mu^2e^x)=-4\mu\cosh\theta$ and we obtain the TBA
equations \eqref{TBA} 
\begin{eqnarray} &&\epsilon_q(\theta)+\frac{1}{2\pi}\int_{-\infty}^\infty d\theta'
\frac{\log(1+e^{-\epsilon_{q-1}(\theta')})}{\cosh(\theta-\theta')}
+\frac{1}{2\pi}\int_{-\infty}^\infty d\theta'
\frac{\log(1+e^{-\epsilon_{q+1}(\theta')})}{\cosh(\theta-\theta')}
\nonumber\\ &&\qquad\qquad\qquad\qquad\mbox{}+D^q=mR\cosh\theta\,\delta_{pq}
\end{eqnarray} where $q=1,2,\ldots,L-2$ and we identify
\begin{equation} 4\mu=mR.
\end{equation}

The Casimir energy is given by the scaling limit of the finite size correction. 
The integral in \eqref{FinSizeIII} is over one period $(L+1)\pi\vareps=2(\log
N-\log\mu)$. For $N$ large, the integrand is of the order $o(1/N)$ unless $y$ is in
one of the scaling regions where $y$ is of the order of $\log N$ or $-\log N$. Hence
\begin{eqnarray}
\log\mf{b}^p(x)&=&\frac{1}{2}
\int_{-(\log N-\log\mu)}^{\log N-\log\mu}\!\!\!dy\,\mf{k}(x-y-\log N)
\log(1+\mf{a}^p(y+\log N))\nonumber\\ &&\quad\mbox{}+\frac{1}{2}
\int_{-(\log N-\log\mu)}^{\log N-\log\mu}\!\!\!dy\,\mf{k}(x+y+\log N)
\log(1+\mf{a}^p(-y-\log N)) +\mf{C}^p\nonumber\\ 
&=&\frac{1}{2\pi N}
\int_{-\infty}^\infty dy\;(e^{x-y}+\mu^2e^{-x+y})\log(1+\hat{a}_+^p(y))\\
&&\mbox{}+\frac{1}{2\pi N}
\int_{-\infty}^\infty dy\; (e^{-x-y}+\mu^2e^{x+y})\log(1+\hat{a}_-^p(y))
+\mf{C}^p+o\left(\frac{1}{N}\right)\nonumber
\end{eqnarray} 
Here we have used the fact that
\begin{equation}
\mf{k}(\pm(x+\log N)) =\frac{e^{-x}+\mu^2e^x}{\pi N} + o\left(\frac{1}{N}\right).
\end{equation} 
For the largest eigenvalue we have $\hat{a}_+(y)=\hat{a}_-(y)$ and
\begin{eqnarray}
\log T_{\text{finite}}^{p,p}(u)&=&\frac{\cosh x}{\pi N}
\int_{-\infty}^\infty dy\;(e^{-y}+\mu^2e^y)\log(1+\hat{a}^p(y))\nonumber\\ 
&=&\frac{2\mu
\cosh x}{\pi N}
\int_{-\infty}^\infty d\theta \cosh\theta \log(1+e^{-\epsilon_p(\theta)})
\\ &=&-\frac{R\sin(L+1)u}{N}\,E_p(R)\nonumber
\end{eqnarray} where the Casimir energy is
\begin{equation} E_p(R)=-\frac{m}{2\pi}\int_{-\infty}^\infty d\theta
\cosh\theta \log(1+e^{-\epsilon_p(\theta)})
\end{equation} and we have used $4\mu=mR$. In the isotropic case $u=\pi/2(L+1)$ and
$\sin(L+1)u=1$.

\section{Regime IV}

In this section we derive the TBA equations for Regime~IV. We use precisely the same
scaling as for Regime~III with $t<0$. It turns out that in this regime, with one
exception when $p=(L-1)/2$, the resulting TBA equations are massless. 

\subsection{Integral Equations and Finite-Size Corrections}

In Regime~IV the analyticity strips are the same as for Regime~III. The leading bulk
contributions to the eigenvalues
$t^{p,q}(u)$ in its analyticity strip are given for large $N$ by
\begin{equation}  t_{\text{bulk}}^{p,q}(u)=
\begin{cases}  t_{\text{const}}^{p,q}\displaystyle{
\left[i\frac{\th_1((L+1)u/2,|t|^{2\nu})}{\th_2((L+1)u/2,|t|^{2\nu})}\right]^N},
&q=p\\ \\  t_{\text{const}}^{p,q}\displaystyle{
\left[\frac{\th_4((L+1)u/2+(L-2p-1)\pi/4,|t|^{2\nu})}
{\th_3((L+1)u/2+(L-2p-1)\pi/4,|t|^{2\nu})}\right]^N}, &q=L-p-1\\ \\
t_{\text{const}}^{p,q},&\mbox{otherwise}
\end{cases}
\end{equation}  where the constants are the same as in Regime~III. Again this has the
required periodicity and zero and pole structure. This applies for $p<(L-1)/2$. For
the case $p>(L-1)/2$, the zeros and poles of order $N$ can be obtained from the
previous case using the relation
\begin{equation} t_0^{p,q}=t_{p'+1}^{p',q},\qquad p'=L-1-p.
\end{equation} In the marginal case $p=p'=(L-1)/2$ we find
\begin{equation}   t_{\text{bulk}}^{p,q}(u)=
\begin{cases}   t_{\text{const}}^{p,q}\displaystyle{
\left[i\frac{\th_1((L+1)u/2,|t|^{\nu})}{\th_2((L+1)u/2,|t|^{\nu})}
\right]^N}, &q=p\\ \\ t_{\text{const}}^{p,q},&\mbox{otherwise.}
\end{cases}
\end{equation} 

As in Regime~III we introduce finite-size correction terms $\ell^q(x)$ by writing
\begin{equation}
\mf{a}^q(x)=t^{p,q}\left(\frac{i}{L+1}x+\frac{p-q}{2}\lambda\right)
=\mf{e}^q(x)\ell^q(x)
\end{equation}  Here, for $p<(L-1)/2$,
\begin{equation}
\mf{e}^q(x)=
\begin{cases} 
\displaystyle{
\left[i\frac{\th_1(ix/2,|t|^{2\nu})}{\th_2(ix/2,|t|^{2\nu})}\right]^N},&q=p\\
\displaystyle{
\left[\frac{\th_4(ix/2,|t|^{2\nu})}{\th_3(ix/2,|t|^{2\nu})}\right]^N},&q=L-p-1\\
1,&\mbox{otherwise}
\end{cases}
\end{equation} and, for $p=(L-1)/2$,
\begin{equation}
\mf{e}^q(x)=
\begin{cases} 
\displaystyle{
\left[i\frac{\th_1(ix/2,|t|^{\nu})}{\th_2(ix/2,|t|^{\nu})}
\right]^N},&q=p\\ 1,&\mbox{otherwise}
\end{cases}
\end{equation}

Following the same Fourier series analysis as in Regime~III and allowing for the change in periodicity
leads to the integral equations
\begin{equation}
\log\mf{a}^q=\log\mf{e}^q+\mf{k}*\log\mf{A}^{q-1}+\mf{k}*\log\mf{A}^{q+1} +\mf{D}^q
\label{TBAIV}
\end{equation}
In this regime the kernel is given by
\begin{eqnarray}
\mf{k}(x)&=&\frac{1}{2(L+1)\pi\vareps}\sum_{k=-\infty}^\infty
\frac{e^{\frac{2ikx}{2(L+1)\vareps}}} {e^{\frac{k\pi}{2(L+1)\vareps}}+
      e^{-\frac{k\pi}{2(L+1)\vareps}}}\nonumber\\
&=&\frac{\th_2(0,\qhat')\th_3(0,\qhat')\th_3(ix,\qhat')}{ 2\pi\,\th_2(ix,\qhat')}
\end{eqnarray}  with the elliptic nomes 
\begin{equation}
\qhat=e^{-\frac{\pi}{2(L+1)\veps}},\qquad
\qhat'=e^{-2\pi(L+1)\veps}=\frac{\mu^4}{N^4}.
\end{equation}  In the critical limit, the kernel again simplifies to
\begin{equation}
\lim_{\veps\to\infty}\mf{k}(x)=\hat{k}(x)=\frac{1}{2\pi\cosh x}.
\end{equation}
The constants $D^q$ can be evaluated as in Regime~III, allowing for the different period, by setting
\begin{eqnarray}
\langle a\rangle=\frac{1}{2(L+1)\pi\vareps}
\int_0^{2(L+1)\pi\vareps} a(x)\,dx
\end{eqnarray}  
Again $D^q$ is a multiple of $\pi i$ with $D^q=0$ for the largest eigenvalue. 

Repeating the analysis of the finite-size corrections for Regime~IV leads to the result
\begin{equation}
\log\mf{b}^p(x)=\int_{-(L+1)\pi\vareps}^{(L+1)\pi\vareps}dy\;
\mf{k}(x-y)\log(1+\mf{a}^p(y))+\mf{C}^p\label{FinSizeIV}
\end{equation}
where $\mf{C}^p$ is a multiple of $\pi i$ that does not contribute to the $1/N$ corrections.
The only difference with Regime~III is in the period.

\subsection{TBA Equations and Casimir Energy}

In the scaling regions we find
\begin{eqnarray}
\log \hat{e}_{\pm}^p&=&\lim_{N\to\infty}\log\mf{e}^p(\pm(x+\log N))\nonumber\\ &=&
\begin{cases} -2e^{-x},&p<(L-1)/2,\quad q=p\\ -2\mu^2 e^x,&p<(L-1)/2,\quad q=L-1-p\\
-2(e^{-x}+\mu^2e^x),&q=p=(L-1)/2.
\end{cases}
\end{eqnarray}
Passing to the scaling limit of the integral equations \eqref{TBAIV} thus yields
\begin{equation}
\log \hat{a}^q=\log \hat{e}^q+\hat{k}*\log \hat{A}^{q-1}+\hat{k}*\log \hat{A}^{q+1}+D^q
\end{equation} 
where again we suppress the subscripts $\pm$. Next, introducing rapidity
variables as before we obtain the TBA equations. For $p=(L-1)/2$ we obtain the same massive TBA
equation as in Regime~III. For $p<(L-1)/2$, however, we obtain the massless TBA equations
\eqref{TBA} 
\begin{eqnarray} &&\epsilon_q(\theta)+\frac{1}{2\pi}\int_{-\infty}^\infty d\theta'
\frac{\log(1+e^{-\epsilon_{q-1}(\theta')})}{\cosh(\theta-\theta')}
+\frac{1}{2\pi}\int_{-\infty}^\infty d\theta'
\frac{\log(1+e^{-\epsilon_{q+1}(\theta')})}{\cosh(\theta-\theta')}
\nonumber\\  &&\qquad\qquad\qquad\qquad\mbox{}+D^q=
\half mR e^{-\theta}\,\delta_{pq}+\half mR e^{\theta}\,\delta_{L-1-p,q}
\end{eqnarray}  where $q=1,2,\ldots,L-2$ and we again identify $4\mu=mR$.

Turning to the finite-size corrections we see that the integral in \eqref{FinSizeIV} is over one
period given by $2(L+1)\pi\vareps=4\log(N/\mu)$.
 Hence
\begin{eqnarray}
\log\mf{b}^p(x)&=&
\int_{-\log (N/\mu)}^{\log (N/\mu)}\!\!\!dy\,\mf{k}(x-y-\log \frac{N}{\mu})
\log(1+\mf{a}^p(y+\log \frac{N}{\mu}))\nonumber\\  &&\mbox{}+
\int_{-\log (N/\mu)}^{\log (N/\mu)}\!\!\!dy\,\mf{k}(x+y+\log \frac{N}{\mu})
\log(1+\mf{a}^p(-y-\log \frac{N}{\mu})) +\mf{C}^p\nonumber\\  &=&\frac{\mu}{\pi N}
\int_{-\infty}^\infty dy\;e^{x-y}\log(1+\hat{a}_+^p(y-\log\mu))\\ &&\mbox{}+\frac{\mu}{\pi
N}
\int_{-\infty}^\infty dy\; e^{-x-y}\log(1+\hat{a}_-^p(y-\log\mu))
+\mf{C}^p+o\left(\frac{1}{N}\right)\nonumber
\end{eqnarray}  Here we have used the fact that
\begin{equation}
\mf{k}(\pm(x+\log \frac{N}{\mu})) =\frac{\mu e^{-x}}{\pi N} +
o\left(\frac{1}{N}\right).
\end{equation}  For the largest eigenvalue we have $\hat{a}_+(y)=\hat{a}_-(y)$ and
\begin{eqnarray}
\log T_{\text{finite}}^{p,p}(u) &=&\frac{2\cosh x}{\pi N}
\int_{-\infty}^\infty dy\;e^{-y}\,\log(1+\hat{a}^p(y))\nonumber\\  &=&\frac{2\mu\cosh
x}{\pi N}
\int_{-\infty}^\infty d\theta e^{-\theta} \log(1+e^{-\epsilon_p(\theta)})
\\ &=&-\frac{R\sin(L+1)u}{N}\,E_p(R)\nonumber
\end{eqnarray} 
Hence the Casimir energy is
\begin{eqnarray}  E_p(R)&=&-\frac{m}{2\pi}\int_{-\infty}^\infty d\theta e^{-\theta}
\log(1+e^{-\epsilon_p(\theta)})\nonumber\\ &=&-\frac{m}{4\pi}\int_{-\infty}^\infty
d\theta \left[ e^{-\theta}\log(1+e^{-\epsilon_p(\theta)})+
e^{\theta}\log(1+e^{-\epsilon_{L-1-p}(\theta)})\right]
\end{eqnarray}  where we have identified $4\mu=mR$ and used the symmetry between
$\epsilon_p(\theta)$ and
$\epsilon_{L-1-p}(-\theta)$. In the marginal case when $p=(L-1)/2$ we see that 
$\epsilon_p(\theta)$ is even and
\begin{equation}   E_p(R)=-\frac{m}{2\pi}\int_{-\infty}^\infty d\theta
\cosh\theta \log(1+e^{-\epsilon_p(\theta)})
\end{equation}  in agreement with the massive case in Regime~III.

\section{Discussion}

In this paper we have given a systematic derivation, at least for the largest eigenvalue, of the TBA
equations for the RSOS lattice models and their fusion hierarchies in the off-critical Regimes~III
and IV related to $A_1^{(1)}$ coset models. Interestingly, in the case of Regime~IV, this appears to
give the first derivation of Zamoldchikov's massless TBA equations describing the renormalization
group flow between distinct coset theories. In a subsequent paper we will similarly, derive the
massive TBA equations for Regimes~I and II pertaining to $Z_{L-1}$ parafermions. In this case the
TBA equations for the largest eigenvalue in Regimes~I and II turn out to be exactly the same due to
a duality between the two regimes. 

The systematic derivation of TBA equations introduced in this paper seems to afford many advantages
over the approach based on Bethe ansatz equations. First, and most importantly, the analysis in this
paper should in principle generalize to allow for the treatment of all the excited states. Second,
the present methods can also be extended to treat systems with a boundary. We hope to explore
these possibilities fully in future work. Already, for the case of $A_4$ in the massless Regime~IV with
fixed boundaries, it is possible~\cite{ChimP} to obtain the TBA equations for all excited states and
to give a complete classification of these eigenvalues. This enables a complete mapping of the flow of
eigenvalues from the tricritical Ising to the critical Ising conformal fixed points.


\section*{Acknowledgements}

This work began while PAP was visiting Amsterdam and Bonn Universities. We thank
Uwe Grimm for help in the early stages of this work. We also
thank Vladimir Bazhanov and Leung Chim for useful discussions. This research is
supported by the Australian Research Council.

\end{document}